# A Fault Location Method Using Direct Convolution: Electromagnetic Time Reversal or Not Reversal

Guanbo Wang, Chijie Zhuang

***Abstract*—Electromagnetic time reversal (EMTR) is drawing increasing interest in short-circuit fault location. In this letter, we investigate the classic EMTR fault location methods and find that it is not necessary to reverse the obtained signal in time which is a standard operation in these methods before injecting it into the network. The effectiveness of EMTR fault location method results from the specific similarity of the transfer functions in the forward and reverse processes. Therefore, we can inject an arbitrary type and length of source in the reverse process to locate the fault. Based on this observation, we propose a new EMTR fault location method using direct convolution. This method is different from the traditional methods, and it only needs to pre-calculate the assumed fault transients for a given network, which can be stored in embedded hardware. The faults can be located efficiently via direct convolution of the signal collected from a fault and the pre-stored calculated transients, even using a fraction of the fault signal.**

## I. INTRODUCTION

It is very important to quickly and accurately locate short-circuit faults in the power system for the continuous and reliable supply of electric power energy. In recent years, EMTR-based methods are being developed to conquer the fault location problem, and have gained extensive interest and research [1]-[10].

The time reversal method was proposed by Bogert of Bell Laboratories, and then applied in the field of electronics [11], radiation sources [12] and electromagnetic waves [13]. Electromagnetic time reversal was first applied to fault location in 2012 [1]. Rachidi et al. have conducted a series of studies on the practicability of EMTR fault location methods [2]-[6]. Furthermore, experiments on EMTR fault location methods are gradually carried out in recent years. Wang et al. conducted fault location simulation experiments in medium voltage distribution networks in China and Switzerland [7][8] respectively. In summary, convincing results have been obtained for EMTR fault location methods both in simulations and experiments.

The EMTR fault location process can be divided into two stages: transient electromagnetic signals caused by faults are collected at the head of the line during the forward process. A series of guessed short-circuit branches are set along the line as the guessed fault position during the reverse process, and the transient signals after the time reversal are injected back into the network. The short-circuit current energy, which is concentrated at the real fault position, is then calculated.

Time reversal is the most important concept for classic EMTR fault location methods. In this letter, we investigate these methods, and show that the time reversal of the collected signals is unnecessary. We find that an arbitrary type and length of source can be injected into the network in the reverse process to locate the fault. On the basis of this observation, we propose a new EMTR fault location method using direct convolution.

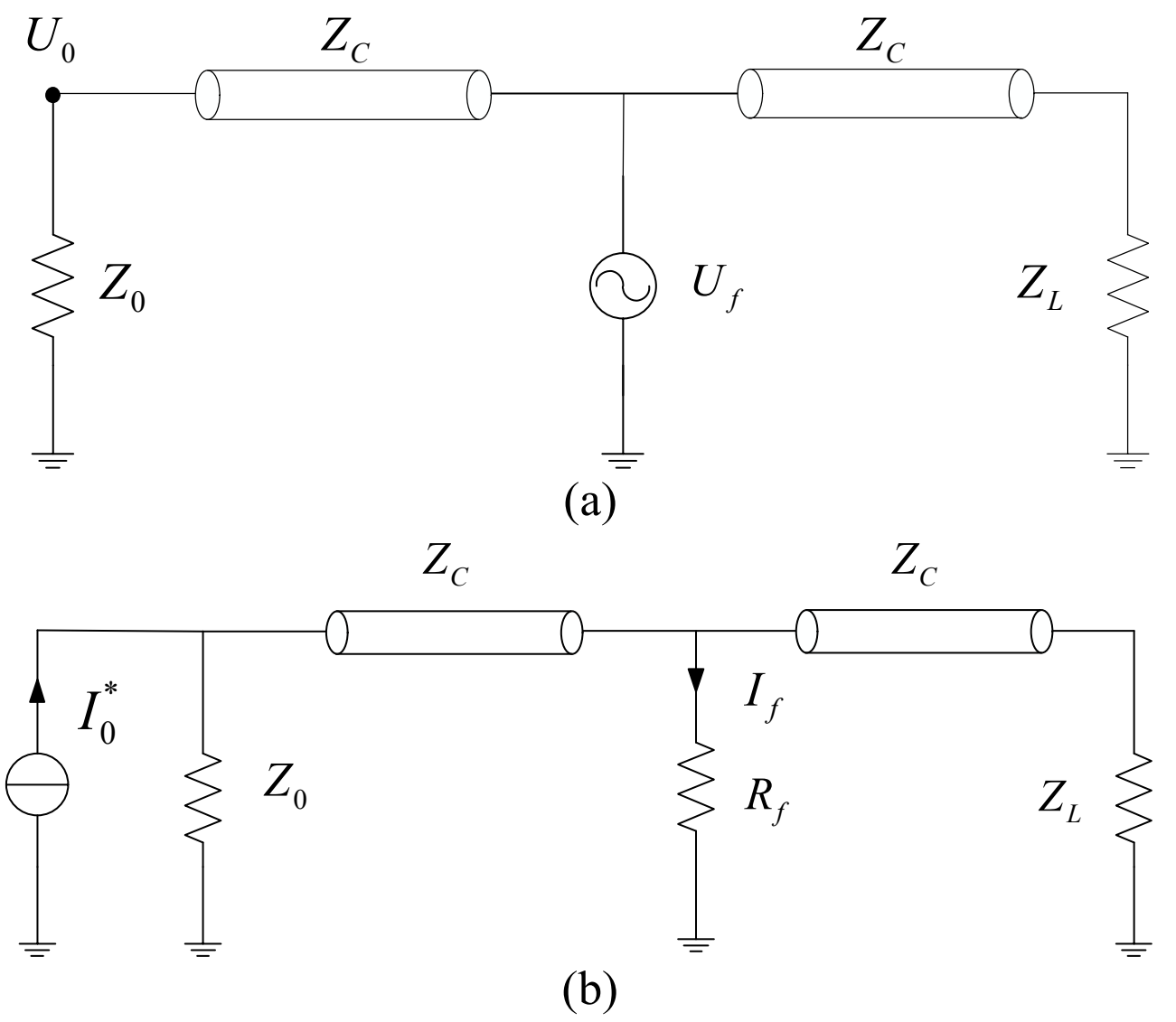

Fig. 1. Representation of a fault along a transmission line and the time reversal process

## II. CLASSIC EMTR FAULT LOCATION METHOD

The basic principle of classical EMTR fault location methods is given in a series of papers [1]-[8]. A short-circuit fault occurs at $x = x_f$ in Fig. 1(a), which is represented by a voltage source $U_f$.

The transient electromagnetic signal generated by the fault is collected at one end of the line (in this case $U_0$ is measured). The expression for $U_0$ in the frequency domain is

$$U_0(\omega) = \frac{(1+\rho_0)e^{-\gamma x_f}}{1+\rho_0 e^{-2\gamma x_f}} U_f(\omega), \quad (1)$$

where $\gamma$ is the line propagation constant without considering loss and is given by $\gamma = j\beta = j\frac{\omega}{c}$ (c is the propagation speed), and $\rho_0$ is the reflection coefficient:

$$\rho_0 = \frac{Z_0 - Z_C}{Z_0 + Z_C} \quad (2)$$

The signal is re-injected into the original system using the Norton equivalent from the same point after the time reversal operation as shown in Fig. 1(b). A short-circuit branch is set at $x = x_f'$ along the line as the guessed fault location. The expression for the short-circuit current in the frequency domain is :

$$I_f(x_f', \omega) = \frac{(1+\rho_0)^2 e^{-\gamma(x_f' - x_f)}}{Z_0(1+\rho_0 e^{-2\gamma x_f'})(1+\rho_0 e^{2\gamma x_f})} U_f^*(\omega) \quad . \quad (3)$$

Assuming that $\gamma$ is purely imaginary, the short-circuit current energy is calculated as:

$$E(x_f^{'}) = \frac{1}{2\pi}\int_{-\infty}^{+\infty} |I_f(x_f^{'},\omega)|^2 \, d\omega = \frac{(1+\rho_0)^2}{2\pi Z_0}\int_{-\infty}^{+\infty} \frac{\left|U_f^*(\omega)\right|^2}{\left|1+\rho_0 e^{-2\gamma x_f^{'}}\right|^2 \left|1+\rho_0 e^{2\gamma x_f}\right|^2} d\omega \quad (4)$$

which reaches the maximum when $x_f' = x_f$. Namely,

$$x_f = \arg\Big|_{x_f^{'}} \max\left(E\left(x_f^{'}\right)\right). \quad (5)$$

From (4) and (5), we note that the extreme point of (4) is irrelevant to the energy-bounded signal $U_f(\omega)$, which will play an important role in the next section.

## III. Time Reversal or Not Reversal

In classic EMTR methods, the transient signal obtained at the terminal of a line is reversed in time during the reverse process.

However, if the signal is injected into the network directly (without time reversal), we can get a short-circuit current in the frequency domain:

$$I_f(x_f^{'},\omega) = \frac{(1+\rho_0)^2 e^{-\gamma(x_f^{'}+x_f)}}{Z_0(1+\rho_0 e^{-2\gamma x_f^{'}})(1+\rho_0 e^{-2\gamma x_f})} U_f(\omega) \quad . \quad (6)$$

The current energy is therefore

$$E(x_f^{'}) = \frac{1}{2\pi}\int_{-\infty}^{+\infty} |I_f(x_f^{'},\omega)|^2 \, d\omega = \frac{(1+\rho_0)^2}{2\pi Z_0}\int_{-\infty}^{+\infty} \frac{\left|U_f(\omega)\right|^2}{\left|1+\rho_0 e^{-2\gamma x_f^{'}}\right|^2 \left|1+\rho_0 e^{-2\gamma x_f}\right|^2} d\omega \quad , \quad (7)$$

which is the same as (4) assuming that $\gamma$ is purely imaginary because $|U_f(\omega)| = |U_f^*(\omega)|$ and $|1+\rho_0 e^{-2\gamma x_f}| = |1+\rho_0 e^{2\gamma x_f}|$. Therefore, it reaches the maximum also at $x_f' = x_f$, which indicates that the time reversal of the transient signal is unnecessary.

For a lossy transmission line, $\gamma$ has a real part,

$$\gamma = \sqrt{(R_0 + j\omega L_0)(G_0 + j\omega C_0)} = \alpha + j\beta \, , \quad (8)$$

where $R_0$, $L_0$, $G_0$ and $C_0$ are the unit parameters of the line. For typical transmission lines, $\alpha$ is very small, e.g., at the order of roughly $10^{-8}$-$10^{-7}$. When $x_f + x_f' = 1000$ km, $e^{-\alpha(x_f'+x_f)}$ is about 0.9-0.99. Therefore, it is reasonable to assume that $\gamma$ is purely imaginary in real applications.

A numerical experiment was carried out on a single transmission line as shown in Fig. 1 to validate the method. The parameters of the line were the same as those of Table II in [9]. A 10-kV power frequency source was applied to the head of the line. Both ends were connected with a large 100-kΩ resistance to serve as the equivalent of power transformers.

The transient signal measured at the head was injected with and without time reversal, Fig. 2 shows the short-circuit current energies at different guessed fault positions. The current energy is almost the same with and without the time reversal of $U_0$, which proves that it is unnecessary to reverse the measured signal.

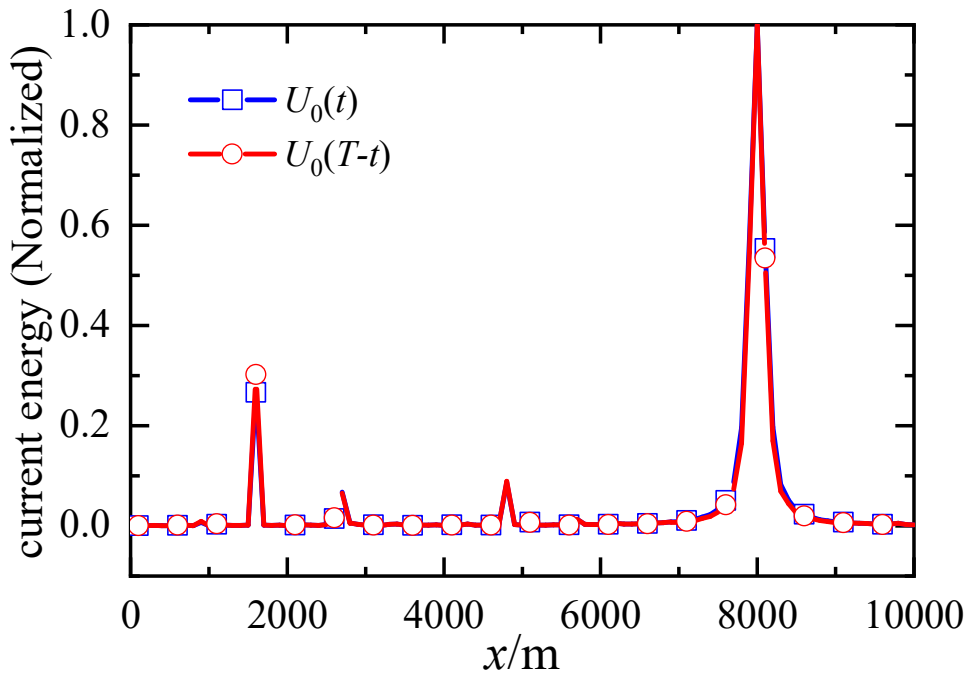

Fig. 2. Normalized current energy at different positions. An ideal short-circuit fault is set at 8 km on the line, and the simulation step size is set to 0.1 μs.

## IV. EMTR Fault Location Method Using Direct Convolution

We have observed that the extreme point of (4) and (7) is irrelevant to the energy-bounded signal $U_f(\omega)$, and is determined only by the denominators of (3) and (6), which include the transfer functions of the reverse and forward processes. Thus we may infer that the reason the energy always reaches the maximum when $x_f' = x_f$ is the specific similarity of the transfer functions of the reverse and forward processes.

Now we assume the voltage signal in (1) is obtained for a fault at the head of the line in the forward process, and an arbitrary energy-bounded signal $U(\omega)$ is injected at the same position in the reverse process. Then the short-circuit current at the guessed fault position is

$$I_f(x_f^{'},\omega) = \frac{(1+\rho_0)e^{-\gamma x_f^{'}}}{Z_0(1+\rho_0 e^{-2\gamma x_f^{'}})} U(\omega) \quad (9)$$

The convolution of the two signals is

$$F(x_f^{'},\omega) = I_f(x_f^{'},\omega)U_0(\omega) = \frac{(1+\rho_0)^2 e^{-\gamma(x_f^{'}+x_f)}}{Z_0(1+\rho_0 e^{-2\gamma x_f^{'}})(1+\rho_0 e^{-2\gamma x_f})} U_f(\omega)U(\omega) \quad (10)$$

Interestingly, (10) has the same structure as (3) and (6), which suggests that it reaches the maximum when $x_f' = x_f$; therefore, we can inject an arbitrary signal in the reverse process to accomplish the fault location.

We now propose a new EMTR fault location method using direct convolution:

(1) [reverse process] For a given network, we can set a series of guessed short-circuit branches along the line, and calculate the short-circuit transient under an arbitrary excitation source at the head of the line. The results are then stored. This can be done before a fault occurs.

(2) [forward process] When a real fault occurs, the fault-generated transient signal is collected at the head of the line.

(3) The transient signal collected in (2) is convoluted with every pre-stored short-circuit transients. The energy of each convoluted signal is calculated, and the case with the maximum energy corresponds to the real fault position.

Table I shows the process of the EMTR fault location method using direct convolution.

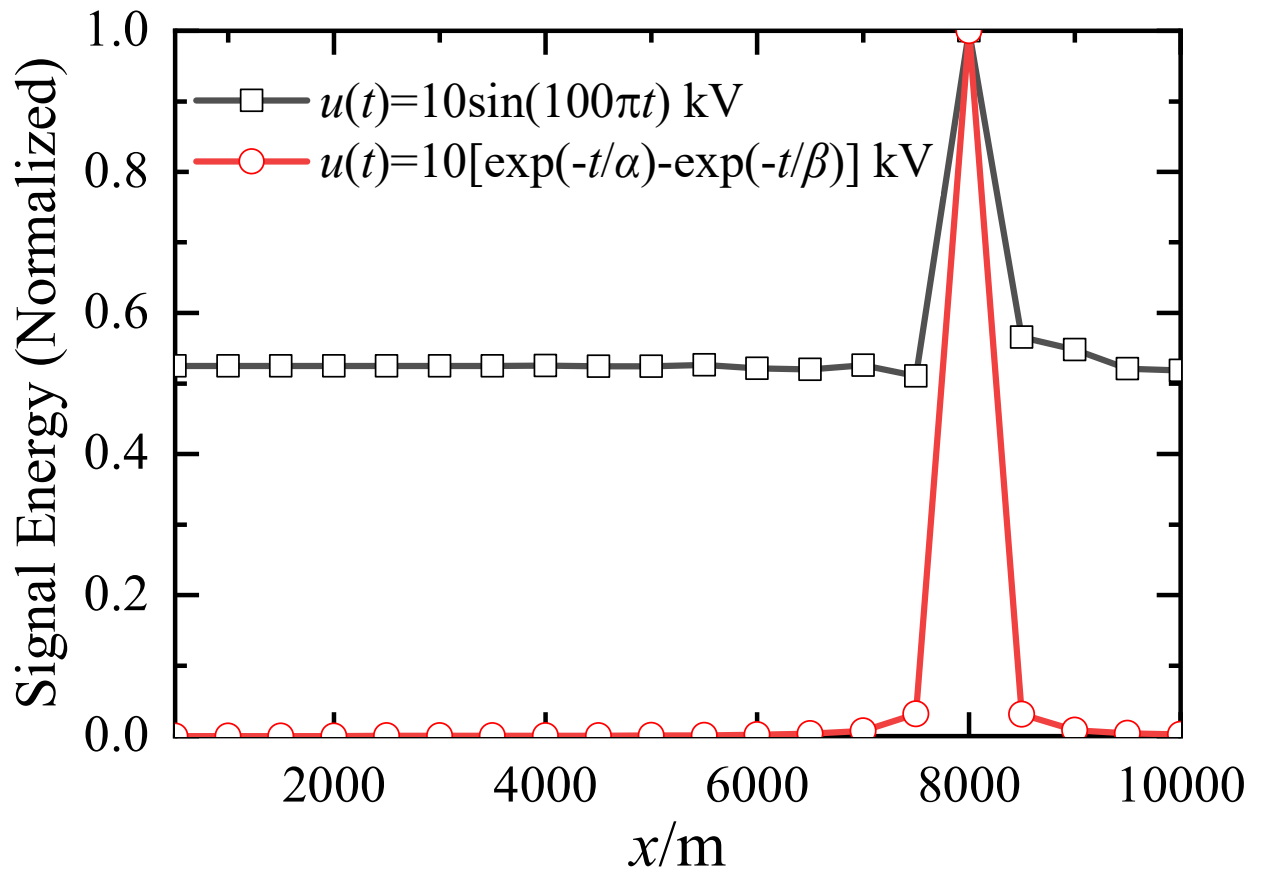

Fig. 3. Normalized energy of $f(t)$ for different $u(t)$ with a power frequency AC source and a lightning impulse. $\alpha$ and $\beta$ equal 20 μs and 3 μs respectively. The sampling length is from 0 to 2 ms.

TABLE. I. PROCESS OF THE EMTR FAULT LOCATION METHOD USING DIRECT CONVOLUTION

| |
|---|
| **Input**: network parameters and topology; an arbitrary excitation signal $u(t)$ (e.g., a lightning impulse); a voltage signal $u_0(t)$ generated by a fault |
| 1. Set a series of short-circuit branches at $x_f'$ as the guessed fault position. 2. $u(t)$ is injected into the network at a terminal of the line. The short-circuit transient $i_f(x_f', t)$ for each short-circuit branch is calculated and stored. 3. When the fault occurs, the transient signal $u_0(t)$ is collected at the same terminal in step 2. 4. Do the convolution of $i_f(x_f', t)$ and $u_0(t)$, e.g., using FFT, which is denoted as $f(x_f', t)$. 5. Calculate the energy of $f(x_f', t)$ which is denoted by $E(x_f')$. |
| **Output**: the predicted fault location is $x_f = \arg\|_{x_f'}\left(\max\left(E(x_f')\right)\right)$. |

To validate the method, different $u(t)$ were chosen to inject into a power line whose parameters were the same as in Section III. The results in Fig. 3 show the fault can be located exactly with different $u(t)$.

In addition, as is shown in Fig. 3, for power frequency signals, the energy at other positions (except the fault position) remains to be relatively high, resulting in a worse contrast ratio; while for the lightning impulse, the energy at other positions stays very low, resulting in a very high contrast ratio. Therefore, high frequency signals like lightning impulses are preferred.

**Because the excitation signal in the reverse process can be arbitrary, the proposed direct convolution method has distinct characteristics**:

(1) For a given network, it only needs to pre-calculate the assumed fault transients once, and the results can be stored, e.g., in embedded hardware, which is different from traditional EMTR fault location methods that require repetitive calculations for each fault. Therefore, in the fault location process, the convolution and energy calculation can be efficiently completed solely in the embedded hardware (e.g., using FFT) without using EMTP software.

(2) The above convolution can be executed using a fraction of the fault signal, and there is no synchronization requirement, as shown later in Section V.

Although equation (10) is given for a single-phase line, we emphasize that the above direct-convolution method works for multiconductor lines; moreover, it also works for the mirrored-minimum energy methods [9][14].

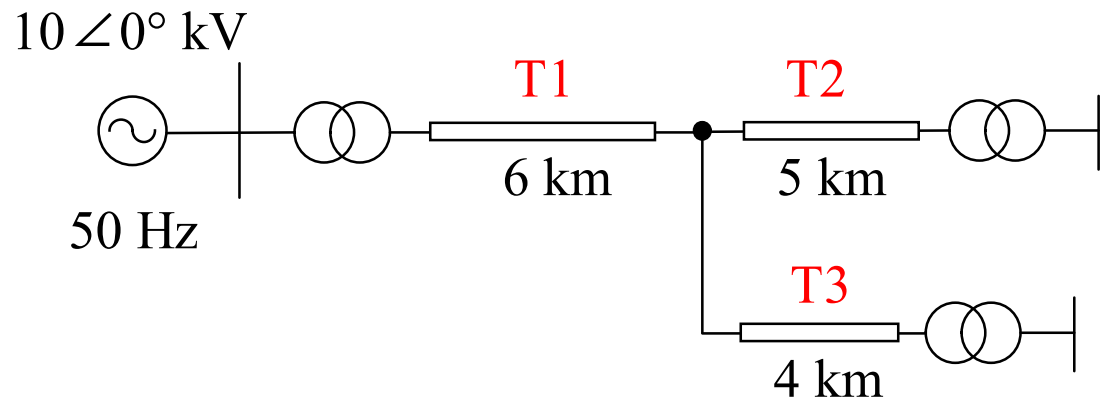

Fig. 4. T-type network model

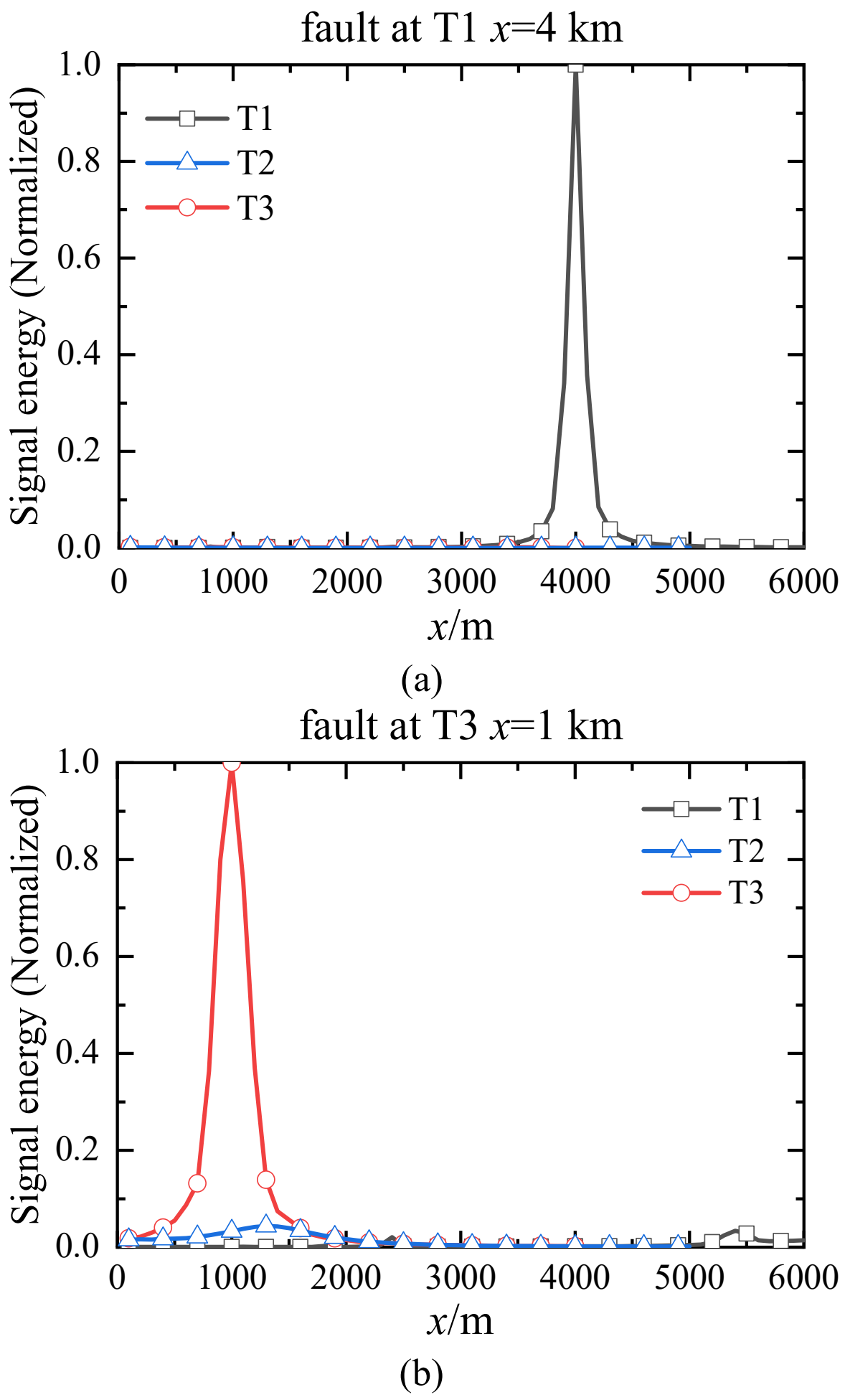

Fig. 5. Normalized signal energy of different fault positions in a T-type network

## V. APPLICATION EXAMPLES

The previous derivation and numerical experiments are for a single transmission line without branches. In this section, we use a T-type power network to study the effectiveness of the method, the influence of signal length, and the ability to use a fraction of the fault transient to locate the fault.

### *A. Application to a T-type network*

We established the T-type power line network shown in Fig. 4. The parameters of the line and transformers remained the same as in Section III. Two faults were set: fault 1 at 4 km on line T1 and fault 2 at 1 km on line T3 along the line respectively, and the fault impedance was set to 1 Ω.

The fault transient signals were collected and convoluted with the response of a lightning impulse. The signal energy reached the maximum at the real fault position as shown in Fig. 5, which illustrates the practicability of the method for T-type networks.

### *B. Convolution using different signal lengths*

Transient signals generated by faults may last for several to tens of milliseconds. We tested the minimum required signal length to locate the fault. For fault 1, we selected the signal with a length from 0.6 ms to 5 ms respectively.

As shown in Fig. 6, the fault can be located correctly when the signal length is 2 ms or more. Therefore, only a fraction of the transient signal is needed.

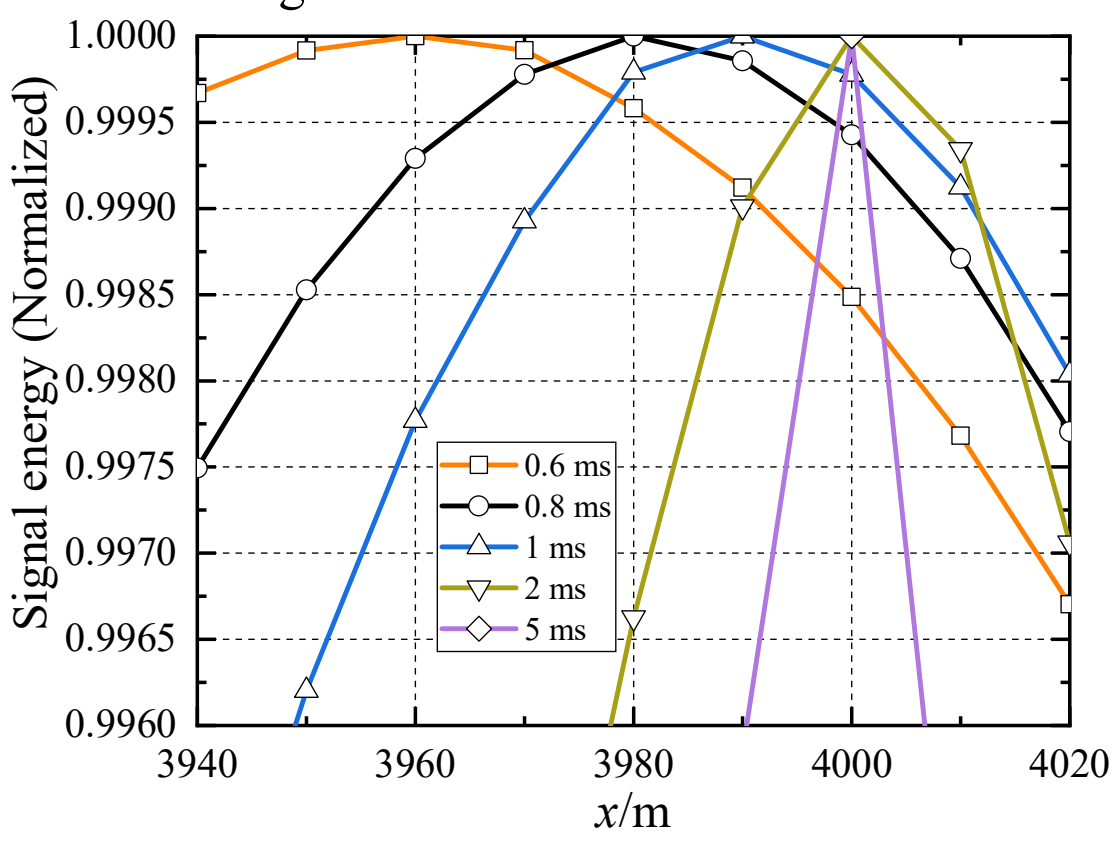

Fig. 6. Normalized signal energies under different signal lengths of 0.6, 0.8, 1, 2 and 5 ms.

We remark that the duration of the injected signal shall be long enough such that the transient wave can reflect in the network for sufficient times, e.g., 20-30 times. Therefore, the minimum length of the transient signal can be estimated using the length of the lines in the network.

### *C. Convolution using an arbitrary continuous fraction of the signal*

Because the injected signal in the reverse process can be arbitrary, we can intercept an arbitrary continuous fraction of the signal to do the convolution. In other words, there is no time synchronization requirement.

We intercepted different fractions with different starting times from the original transient signal to locate fault 1.

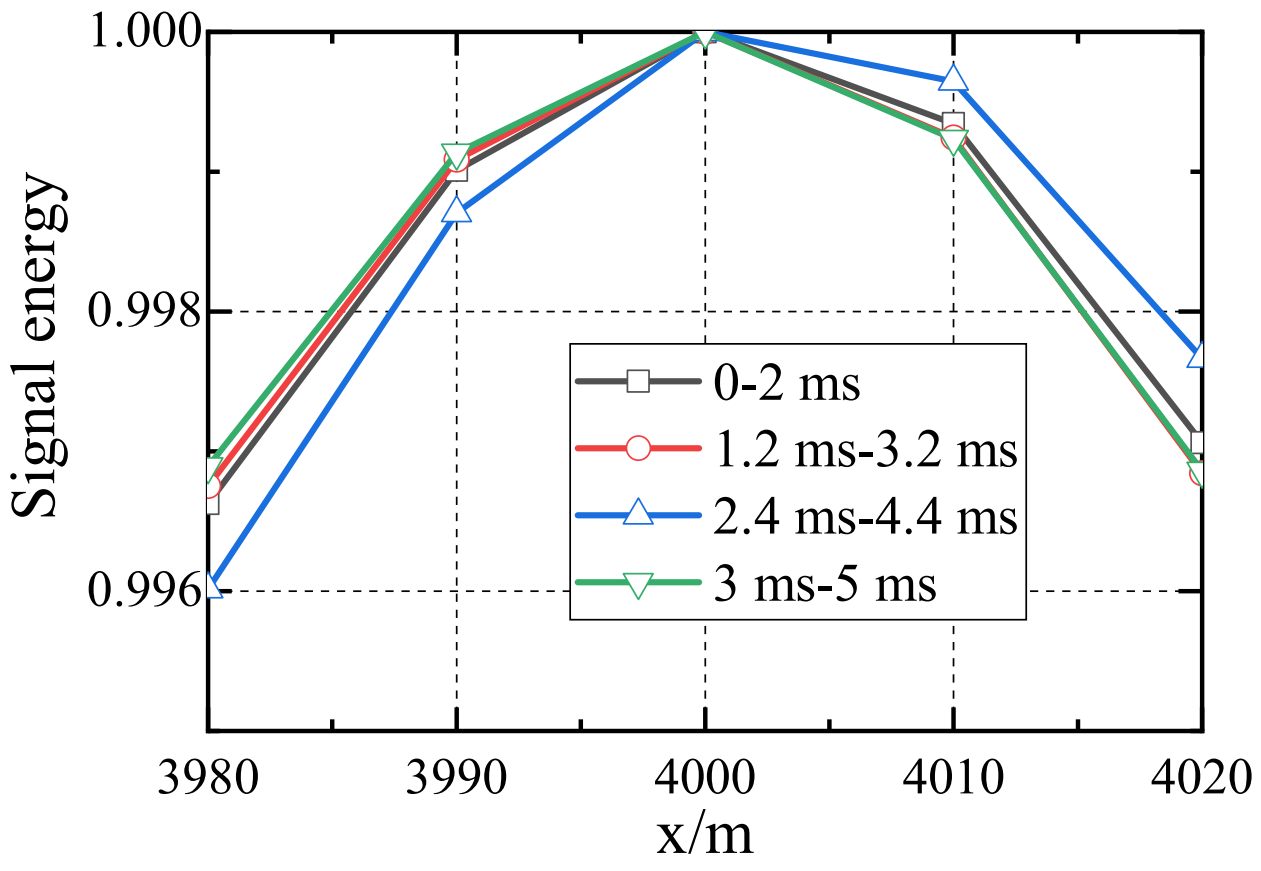

Fig. 7. Normalized signal energies of different fractions of the transient signal generated by fault 1.

Fig. 7 shows that the method correctly locates the fault independent of the starting time of the transient signal. Therefore, it is not necessary to synchronize a given transient signal with the pre-computed signal. The location can be accomplished as long as we intercept data of the same length when doing the convolution.

## VI. Conclusion

We have investigated the classic EMTR fault location methods, and found that it is not necessary to reverse the obtained signal in the reverse process as is done in these methods. The effectiveness of EMTR fault location methods results from the specific similarity of the transfer functions in the forward and reverse processes. Therefore, we can inject an arbitrary type and length of source in the reverse process to locate the fault.

On the basis of above observation, we proposed a new EMTR fault location method using direct convolution. For a given line, we can set a series of short-circuit branches along the line, and pre-calculate the short-circuit transient under an arbitrary excitation source such as a lightning impulse; the results are then stored. When a real fault occurs, the fault-generated transient signal is collected and convoluted with every stored short-circuit transients. The energy of each signal obtained from the convolution is calculated, and the case with the maximum energy corresponds to the real fault position.

The proposed method is different from the traditional EMTR fault location method because it only needs to use EMTP software once to pre-calculate the assumed fault transients and the results can be stored in such media as embedded hardware. In the fault location process, the convolution and energy calculation can be efficiently completed solely in the embedded hardware.

We emphasize that this convolution can be executed using a fraction of the fault signal, and there is no synchronization requirement.

Numerical experiments were carried out on a T-type power network that showed the effectiveness of the method.